\documentclass[10pt]{iopart}
\usepackage{iopams}  
\expandafter\let\csname equation*\endcsname\relax
\expandafter\let\csname endequation*\endcsname\relax
\usepackage{amsmath, amsthm, amssymb}
\usepackage{graphicx} 
\usepackage[leftcaption]{./sidecap}
\usepackage{epsfig}
\usepackage{dcolumn}    
\usepackage{xspace} 
\usepackage{slantsc}

\newcommand{\vek}[1]{\boldsymbol{#1}}

\begin{document}

\title[On precessing convention]{Post-Newtonian analysis of precessing convention for spinning compact binaries}

\author{A Gupta$^{1,2}$ and A Gopakumar$^{1}$}
\address{$^1$Department of Astronomy and Astrophysics, Tata Institute of Fundamental Research, Mumbai 400005,
India\\
$^2$Inter University Centre for Astronomy and Astrophysics, Ganeshkhind, Pune 411007, India}
\ead{anuradha@iucaa.ernet.in, gopu@tifr.res.in}

\begin{abstract}
A precessing source frame, constructed using the Newtonian orbital angular momentum $\vek L_{\rm N}$,
can be invoked to model inspiral gravitational waves from generic spinning compact binaries.
An attractive feature of such a precessing convention is its ability to remove all spin precession induced
modulations from the orbital phase evolution. 
However, this convention usually employs a
post-Newtonian (PN) accurate precessional equation, appropriate for the PN accurate orbital angular momentum $\vek L$,
to evolve the $\vek L_{\rm N}$-based precessing source frame.
This influenced us to develop 
inspiral waveforms for spinning compact binaries in a precessing convention
that explicitly employ $\vek L$ to describe the binary orbits.
Our approach introduces 
certain additional 3PN order terms in the evolution equations for the orbital 
phase and frequency with respect to the usual $\vek L_{\rm N}$-based 
implementation of the precessing convention.
We examine the practical implications of
these additional terms  by computing the match
between inspiral waveforms that employ  $\vek L$ and $\vek L_{\rm N}$-based 
precessing conventions. 
The match estimates are found to be smaller than the optimal value, namely $0.97 $, for 
a non-negligible fraction of unequal mass spinning compact binaries.

\end{abstract}


\pacs{04.25.Nx, 04.30.-w, 97.60.Lf, 95.30.Sf}
\submitto{\CQG}


\section{Introduction}

 Inspiralling compact binaries containing spinning neutron stars and (or) black holes (BHs) 
are key sources for the network of second generation interferometric gravitational wave (GW) detectors
\cite{SS_lr}.
These instruments include the two advanced LIGO (aLIGO) observatories \cite{Harry10}, the advanced Virgo \cite{virgo2}, the KAGRA \cite{KS11},
the GEO-HF \cite{GEO2} and the planned LIGO-India \cite{LIGO_I_Unni}.
In contrast, massive spinning BH binaries are one of the most exciting sources for the space-based GW observatory 
like the planned eLISA \cite{eLISA}.
GWs from such inspiralling compact binaries, whose components are specified by their masses and spins,
can be accurately modeled using perturbative approaches to tackle the underlying Einstein field equations \cite{LR_LB}.
Therefore, the optimal detection technique of {\it matched filtering} can be employed to detect and 
characterize inspiral GWs from such binaries \cite{JK_LRR,Vitale14}.
This involves cross correlating the interferometric output data with a bank of
templates that theoretically model inspiral GWs from spinning compact binaries.
A successful detection demands that at least one template should remain in phase 
as much as possible  with the buried weak GW signals in the frequency windows of 
various GW observatories \cite{DIS98}.

  During the GW emission induced inspiral, dynamics and associated GWs from compact binaries can be accurately 
described using the post-Newtonian (PN) approximation to general relativity \cite{LR_LB,Will}.
The PN description provides interesting quantities, required for various template constructions, as an 
asymptotic series in terms of certain dimensionless parameter.
For binaries in quasi-circular orbits, it is usual to use  
$x=(G\, m\, \omega/c^3)^{2/3}$ as the PN expansion parameter while constructing
inspiral templates, where $m$ and $\omega$ stand 
for the total mass and orbital (angular) frequency of the binary \cite{BDI,Boyle_07}.
Currently, GW frequency and associated phase evolution,  crucial inputs to construct various inspiral 
template families, are 
known to 3.5PN order
for non-spinning compact binaries \cite{BFIJ,BDFI}.
In other words,  PN corrections to the above two quantities are computed to the $x^{7/2}$ order 
beyond the leading quadrupolar (Newtonian) order for such binaries.
Moreover, the amplitudes of 
the two GW polarization states, $h_{\times}$ and $ h_+$, for non-spinning binaries are available to 3PN order \cite{BF3PN}.
Very recently, detailed computations led to the determination of 
the dynamics of such binaries to the 4PN order \cite{4PN}.  
Binaries that contain  compact objects with intrinsic rotations, the spin effects enter 
the dynamics and GW emission via spin-orbit and spin-spin interactions \cite{BO_75, Tulczyjew}. 
In binaries containing maximally spinning BHs, the spin-orbit coupling (linear in
the spins) first appears at the 1.5PN order, while the spin-spin interaction (which is
quadratic in spins) first occurs at the 2PN order \cite{LK_95}.
Additionally, $\vek S_1$, $\vek S_2$ and $\vek L$, the two spin and orbital angular momenta,
for generic spinning compact binaries precess around the total angular momentum 
$\vek J=\vek L+\vek S_1+\vek S_2$ due to spin-orbit and  spin-spin interactions.
This forces substantial modulations of the emitted GWs from inspiralling generic spinning 
compact binaries compared to their non-spinning counterparts \cite{LK_95, ACST}.
Therefore, it is important to incorporate various effects due to the intrinsic rotations 
of compact objects while constructing inspiral GW templates for spinning compact binaries.
At present, GW frequency evolution and amplitudes of $h_{\times}$ and $ h_+$ for BH binaries 
having maximally spinning components are fully determined to 2.5PN and 2PN orders, respectively,  
while incorporating all the relevant spin induced effects \cite{LK_95,ABFO,BBF, Alvi, FBB,BFH}.
Moreover, the on-going detailed computations are providing various higher PN order spin-orbit 
and spin-spin contributions to the dynamics of spinning compact binaries in general orbits 
and to the orbital frequency evolution for quasi-circular inspiral.
At present, the spin-orbit contributions to binary dynamics and GW frequency
evolution are available up to the next-to-next-to-leading order (2PN order) 
beyond the leading order \cite{Bohe2013, 2PN_SO} while adapting the MPM (Multipolar post-Minkowskian) approach
\cite{LB_2009}. In contrast, the higher order spin-spin contributions to the binary dynamics
are usually tackled in the Arnowitt-Deser-Misner canonical formalism \cite{GS_2009}
and in the Effective Field Theory formalism \cite{GR_04, Porto2006} 
(Note that the spin-orbit effects in the Effective Field Theory formalism are computed, for example, in \cite{EFT_SO}).
These approaches provide various spin(1)-spin(2) and  spin-squared 
contributions to the orbital dynamics \cite{GS_group}. Moreover, various source multipole moments needed to obtain the
spin contributions to GW luminosity at the 3PN order and GW polarization states to the 2.5PN order were computed in \cite{porto}.

   The construction of time-domain $h_{\times}$ and $h_+$ associated with inspiralling generic spinning compact binaries
requires us to numerically solve a set of PN accurate differential equations for 
$\vek S_1, \vek S_2,\vek L_{\rm N}, x $ and the orbital phase \cite{LK_95}, where $\vek L_{\rm N}$ is the
Newtonian orbital angular momentum. 
The numerical integration provides temporal evolutions for the orbital phase, the associated angular frequency and the two angles
that specify the orientation of the orbital plane in an inertial frame associated with the direction of 
$\vek J$ at the initial epoch. 
These variations are incorporated into the PN accurate expressions for $h_{\times}$ and $ h_+$ to obtain
PN accurate time-domain inspiral waveforms from such binaries \cite{LK_95}. In this approach,
the differential equation for the orbital phase explicitly depends on the precessional motion of the orbital
plane \cite{LK_95, ACST}. Therefore, it is not possible to express 
the orbital phase 
as an integral of the orbital frequency as usually done in the case of non-spinning compact binaries \cite{DIS98}.
A decade ago, Buonanno, Chen and Vallisneri proposed an approach, referred to as the precessing convention,
that factorizes the generic spinning binary waveform into  a {\it carrier signal} and a {\it modulated amplitude} term
 \cite{BCV}.
In this approach, the phase of the carrier signal ($\Phi_p$) essentially 
coincides with the accumulated orbital phase such that $\dot {\Phi}_{p} \equiv \omega$.
Moreover,  the precessional dynamics of the orbital plane only influences
the modulated amplitude part of inspiral waveform even for generic compact binaries.
Therefore, the precessing convention  
disentangles  the precessional effects from its non-precessional counterparts while modeling 
both the amplitude and the phase evolutions of inspiral GWs from such astrophysical systems. 
This convention was employed to model inspiral GW signals from compact binaries containing misaligned
single-spin and to probe its data analysis benefits \cite{PBCV04, BCPTV05}.
Very recently, inspiral-merger-ringdown waveforms for generic spinning BH binaries, invoking the effective-one-body
approach \cite{TD}, also adapted the precessing convention to model GWs from the inspiral part \cite{Pan_etal_14}.
We note that this convention requires a {\it precessing source frame} which is usually 
based on the Newtonian orbital angular momentum $\vek L_{\rm N}$.  
However, \cite{Pan_etal_14} employed both $\vek L $ and $\vek L_{\rm N}$ to  model GWs during the late part of 
the binary inspiral just prior to the plunge.

 In this paper, we develop 
 a prescription to compute PN accurate inspiral waveforms for generic spinning 
compact binaries while using the PN accurate orbital angular momentum
$\vek L $ to construct the precessing source frame. 
This is influenced by the usual practice of employing precessional equation appropriate for 
 $\vek L$  
to evolve $\vek L_{\rm N}$
and the associated precessing source frame 
while constructing inspiral waveforms via the precessing convention of \cite{BCV}.
We show that the use of such an adiabatic approximation, namely employing 
an orbital averaged differential equation for $\vek L_{\rm N}$,
can lead to PN corrections to  $\dot {\Phi}_{p} = \omega$.
These observations motivated us to provide a set of PN accurate equations to obtain temporally evolving 
quadrupolar order inspiral GW polarization states for generic spinning compact binaries
in the $\vek L $-based precessing convention. In our approach, 
the spin precession induced modulations enter the differential equation for the orbital phase
only at the 3PN order. Moreover, the $\vek L $-based convention 
requires us to include additional 3PN order terms in the differential equation for $x$, compared
to the usual $\vek L_{\rm N}$-based approach.
We explore the practical 
implications of these additional terms with the help of {\it match} computations, detailed in \cite{DIS98, BO96}.
The match computations involve two inspiral families where one is constructed 
via our $\vek L$-based precessing convention and therefore incorporate the above mentioned 3PN order terms.
The other family is based on 
the {\scshape lalsuite} SpinTaylorT4 code, developed by the LIGO Scientific Collaboration (LSC) \cite{LAL},
that implemented the $\vek L_{\rm N}$-based precessing convention of \cite{BCV}.
In this code, the precessional dynamics is fully 2PN accurate while 
the differential equation for $x$ incorporates
 spin-orbit contributions to 3.5PN order. For match computations,
we employ PN accurate relation connecting $\vek L_{\rm N}$ and $\vek L$ to construct two waveform
families with physically identical initial orbital and spin orientations.
We terminate the two inspiral waveform families when their respective
$x$ parameter reaches $0.1$ which roughly corresponds to orbital separations $\sim10\,G\,m/c^2$.
These considerations allow us to attribute the reported match estimates to the 
above mentioned additional 3PN order
terms present in our  differential equations for the orbital phase and frequency.
We find that the match estimates are less than the optimal 0.97 value for a non-negligible fraction 
of unequal mass spinning compact binaries. It may be recalled that 
such an optimal match value
 roughly corresponds to a $10\%$ loss in the ideal event rate.

   In what follows, we briefly summarize the usual implementation of the $\vek L_{\rm N}$-based precessing convention 
   and explore the consequence of employing an orbital averaged precessional equation for  $\vek L_{\rm N}$.
In section~\ref{Sec_L}, we detail the construction of 
 quadrupolar order GW polarization states in our $\vek L$-based precessing convention.
 The match estimates involving these two inspiral families with physically equivalent 
initial configurations 
and associated discussions are listed in 
section~\ref{result} while section~\ref{Sec_dis} provides
a brief summary.

\section{Inspiral waveforms in $\vek L_{\rm N}$ and $\vek L$-based  precessing conventions }

 We begin by summarizing the usual implementation of the $\vek L_{\rm N}$-based precessing convention 
and explore the  consequence of employing an orbital averaged differential equation for $\vek L_{\rm N}$.

\subsection{The $\vek L_{\rm N}$-based precessing convention }
\label{SecII_A}
   The precessing convention, introduced in \cite{BCV}, aims to 
remove all the spin precession induced
modulations from the orbital phase evolution. 
In this approach, the orbital phase $\Phi_p(t)$ is written 
as an integral of the orbital frequency $\omega(t)$, namely $\Phi_p(t)=\int \omega(t')\, dt'$, even for 
generic spinning compact binaries.
This feature is crucial to ensure that 
the inspiral waveform for a precessing spinning  compact binary can be written 
as the product of a non-precessing carrier waveform and a modulation term that contains all the precessional effects.
This is how the approach  disentangles the precessional effects from
their non-precessional counterparts both in the amplitude and phase of inspiral waveforms.
It should be noted that 
 in the absence of precessing convention, the orbital phase of a generic spinning binary is given by
$\int [ \omega(t') -\dot{\alpha}'(t')\,\cos \iota'(t') ] \, dt'$, where $\iota'$ and $\alpha'$ specify the orientation of $\vek L_{\rm N}$
in an inertial frame associated with the initial direction of $\vek J$ \cite{LK_95}.

   To obtain inspiral waveforms for spinning compact binaries in their precessing convention, \cite{BCV} employed 
certain {\it precessing source frame} 
$( \vek e_1^l, \vek e_2^l, \vek e_3^l \equiv \vek l)$,
where $\vek l$ is the unit vector along $\vek L_{\rm N}$.
The basis vectors of this triad satisfy the evolution equations
$\dot{\vek e}^l_{1,2}=\vek \Omega^l_e \times \vek e^l_{1, 2}$ and 
$\dot{\vek e}^l_{3}\equiv \dot{\vek l}=\vek \Omega_{k} \times \vek l$.
The angular frequency $\vek \Omega^l_e$ is constructed in such a manner that these
 three basis vectors always form an orthonormal triad. 
This is possible with the following expression for 
$\vek \Omega^l_e$, namely  
$\vek \Omega^l_e= \vek \Omega_k - (\vek \Omega_k \cdot \vek l)\vek l$,
where $\Omega_k$ is the usually employed precessional frequency for $\vek l$. 
The expression for $\vek \Omega_{k}$ that includes the dominant order spin-orbit 
and spin-spin contributions can be obtained
by collecting the terms that multiply $\hat {\vek L}_{\rm N}$ $(\equiv \vek l)$ in equation (9) of \cite{BCV}.
With the help of $\vek e_1^l $ and $\vek e_2^l$, 
an orbital phase $\Phi_p(t)$  may be defined such that  $\vek n = \cos \Phi_p \, \vek e^l_1 +\sin \Phi_p \,\vek e^l_2$,
where $\vek n$ is the unit vector along the binary separation vector $\vek r$. 
Additionally, one 
 may define a co-moving frame ($\vek n, \vek \lambda = \vek l \times \vek n, \vek l$)
such that the time derivative of $\vek n$ is given by 
$\dot{\vek n } = \dot{\Phi}_p \vek \lambda + \vek \Omega^l_e \times \vek n$.
It was argued in \cite{BCV} that $\vek \Omega^l_e$ should only be proportional to $\vek n$ which ensures that 
$\dot{\vek n } \cdot \vek \lambda = \dot{\Phi}_p$.
This leads to the desirable expression for $\dot{\Phi}_p$, namely $\dot{\Phi}_p=\omega$, while employing the 
adiabatic condition for the sequence of circular orbits: $\dot{\vek n } \cdot \vek \lambda = \omega$.
This adiabatic condition is equivalent to $\dot{\vek n} \cdot \dot{\vek n} \equiv  \omega^2$
that provides another version of the PN independent relation connecting the 
linear and orbital angular velocities, $ v^2 \equiv r^2\,\omega^2$. 

   
 A close inspection reveals that 
this approach usually employs an orbital averaged differential equation for  $\vek L_{\rm N}$
to evolve the precessing source frame while constructing PN accurate inspiral waveforms.
 It turns out that the differential equation for  $\vek L_{\rm N}$ in such an 
 adiabatic approximation is identical to the evolution equation for 
 the PN accurate orbital angular momentum $\vek L$ \cite{GG1,GS11}.
In what follows, we explore the effect of 
 such an adiabatic approximation  
 on the equation for $\dot {\vek n}$ in the ($\vek n, \vek \lambda, \vek l$)
co-moving frame and on the derivation of $\dot{\Phi}_p$ equation.
The usually employed expression for $\vek \Omega_k$
 while considering only the leading order spin-orbit interactions may be written as
\begin{align}
\label{Eq_l_omega_k}
\vek \Omega_k &=  \frac{c^3}{Gm}\, x^3\,\Bigg\{ \delta_1\,q\, \chi_1\, \vek s_1  
+\frac{\delta_2}{q}\, \chi_2\,  \vek s_2  \Bigg\} \,,
\end{align}
where $q=m_1/m_2$ ($m_1 \geq m_2$) is the mass ratio and $\delta_{1,2} = \eta/2 + 3\,(1\mp \sqrt{1-4\eta})/4$ while 
 $\eta=m_1\,m_2/m^2$ is 
the symmetric mass ratio. 
The Kerr parameters $\chi_1$ and $\chi_2$ of the two compact objects of mass $m_1$ and $m_2$ specify 
their spin angular momenta by $\vek S_{1,2}=G\, m_{1,2}^2\, \chi_{1,2}\,\vek s_{1,2}/c$, 
where $\vek s_1$ and $\vek s_2$ are the unit vectors along $\vek S_1$ and $\vek S_2$.
It is straightforward to show that the above equation is identical to $\omega^2$ terms that multiply
$\hat { \bf L}_{\rm N}$ on the right hand side of equation (9) in \cite{BCV}.
To explore the implication of using the above expression for $\vek \Omega_k$ to 
construct $\vek \Omega^l_e $, 
we revisit the arguments detailed in the appendix~B of \cite{BCV}.
These arguments, crucial to obtain  $\dot{\Phi}_p=\omega$,
require that 
 $\vek \Omega^l_e \times \vek n$, appearing in the equation
$\dot{\vek n } = \dot{\Phi}_p \vek \lambda + \vek \Omega^l_e \times \vek n$, should be zero.
A closer look reveals that \cite{BCV} did not use the explicit expression for 
$\vek \Omega^l_e$ to show that $\vek \Omega^l_e$ lies along $\vek n$.
Instead, the authors arrived at such a conclusion 
with the help of the following two steps.
First, it was noted that 
$ \dot {\vek l}$ should be 
$\propto \vek \lambda$ (see lines around  equations~(B3), (B4) and (B5) in \cite{BCV}).
This was inferred by employing  
the definition for $\vek l$ ($\vek l = \vek n \times \vek \lambda$),
the resulting time derivative for $\vek l$ ($ \dot {\vek l} = \dot {\vek n} \times \vek \lambda + \vek n \times \dot { \vek \lambda}$),
the adiabatic condition for circular orbits ($\dot {\vek n} = \omega \,\vek \lambda)$ and 
the time derivative for $\vek \lambda$ in the co-moving frame ($\dot {\vek \lambda} = - \dot{\Phi}_p \, \vek n + 
\vek \Omega^l_e \times \vek \lambda $).
In the second step, \cite{BCV} invoked  
the requirement that $ \dot {\vek l}$ should also be given by  $\vek \Omega^l_e \times \vek l$
in the precessing source frame.
With the help of the above two arguments, namely $ \dot {\vek l} \propto \vek \lambda $  and 
$ \dot {\vek l}= \vek \Omega^l_e \times \vek l$, \cite{BCV} concluded that 
$\vek \Omega^l_e$ should lie along $\vek n$ as $\vek \lambda = \vek l \times \vek n$.
We would like to emphasize that \cite{BCV} never invoked their explicit expression for $\vek \Omega^l_e$
to show that $\vek \Omega^l_e$ can  have components only along $\vek n$ which is essential to 
obtain the relation $\dot{\Phi}_p=\omega$.
In fact, it is straightforward to show with the help of equation~(\ref{Eq_l_omega_k}) that
$\vek \Omega^l_e$ can have components along $\vek \lambda$ since
$\vek \Omega^l_e \cdot \vek \lambda = (c^3/Gm)\, x^3\,( \delta_1\,q\, \chi_1\, \vek s_1 \cdot \vek \lambda  
+\delta_2 \, \chi_2/q  \,\vek s_2 \cdot \vek \lambda) \neq 0$,  
in general.
This results in the following 1.5PN accurate expression for $ \dot {\vek n}$
\begin{align}
\label{Eq_ndot}
\dot {\vek n } &=  \dot{\Phi}_p \vek \lambda +  \frac{c^3}{Gm}\, x^3\,\Bigg\{  \delta_1\,q\, \chi_1\,  \bigl [\vek s_1 \times \vek n - (\vek s_1 \cdot \vek l)\, \vek \lambda \bigr] 
+\frac{\delta_2}{q}\, \chi_2\, \bigl [\vek s_2 \times \vek n - (\vek s_2 \cdot \vek l)\, \vek \lambda \bigr]   \Bigg\} \,.
\end{align}
Clearly, the 1.5PN order terms that arise from $\vek \Omega^l_e \times \vek n$ in the above equation will not be zero 
for generic spinning compact binaries.

Interestingly,  it is still possible to equate 
$\dot{\Phi}_p$ to $ \omega$ by employing the adiabatic condition $\dot{\vek n } \cdot \vek \lambda = \omega$
as $( \vek \Omega^l_e \times \vek n ) \cdot \vek \lambda =0$ even 
in the presence of non-vanishing 1.5PN order $\vek \Omega^l_e \times \vek n$ term.
However, the equivalent version of the adiabatic condition, namely $\dot{\vek n} \cdot \dot{\vek n} \equiv  \omega^2$,
forces the  differential equation for $\Phi_p$ to contain 3PN order corrections in addition to 
the standard $\omega$ term. 
In other words, two equivalent versions of the same adiabatic condition for circular orbits, namely
$\dot{\vek n } \cdot \vek \lambda = \omega$ and  $\dot{\vek n} \cdot \dot{\vek n} \equiv  \omega^2$,
provide different evolution equations for $\Phi_p$.
This is the unexpected consequence of 
employing precessional equation appropriate for $\vek L$ to evolve the $\vek L_{\rm N}$-based precessing source frame.
It is not difficult to deduce that 
this essentially arises from the non-vanishing 1.5PN order $\vek \Omega^l_e \times \vek n$ contributions to $ \dot {\vek n}$
listed in the above equation. 
The following arguments can also be used to clarify why the above two versions of the same adiabatic condition for the 
circular orbits
result in two different expressions for $ \dot{\Phi}_p$.
In our opinion, this arises by identifying $\vek k \times \vek n$ to be  $\vek \lambda =\vek l \times \vek n  $, where 
 $\vek k$ is the unit vector along 
$\vek L$. 
Strictly speaking, the precessing source frame of \cite{BCV} is based on $\vek L$ rather than $\vek L_{\rm N}$
due to the use of $\vek \Omega_k$ that provides the  precessional equation for $\vek L$ \cite{GG1, GS11}.
This implies that their co-moving triad is rather $\vek L$-based 
and the expression for $ \dot {\vek n} $ got components along $\vek k \times \vek n$ instead of $\vek \lambda = \vek l \times \vek n$.  
Therefore,  $\dot{\vek n } \cdot \vek \lambda $ results in PN corrections to 
$ \dot{\Phi}_p$ as $ (\vek k \times \vek n)  \cdot \vek \lambda$
is unity only at the leading order (this may be deduced 
 from our equation~(\ref{Eq_l_k}) listed below).
This leads to a differential equation for $ \dot{\Phi}_p$ that involves PN corrections to $\omega$.

The use of precessional equation appropriate for $\vek L$ while implementing the 
 precessing convention of \cite{BCV} motivated us to 
develop a $\vek k$-based precessing convention for constructing inspiral
waveforms for spinning compact binaries.
This should also allow to explore the practical implications of using 
the adiabatic approximation to evolve $\vek L_{\rm N}$ in the 
usual implementation of the precessing convention.


\subsection{Inspiral waveforms via an $\vek L $-based precessing convention }
\label{Sec_L}
 
  Influenced by the above arguments and \cite{BCV}, we first introduce a $\vek k$-based precessing source frame: 
($\vek e_1$,$\vek e_2$, $\vek e_3 \equiv \vek k$).
The precessional dynamics of $\vek e_1, \vek e_2$ and $\vek e_3$ are provided by 
$\dot {\vek e}_{1,2,3}=\vek \Omega_e \times \vek e_{1,2,3}$, where 
$ \vek \Omega_e \equiv \vek \Omega_k -  ( \vek \Omega_k \cdot \vek k)\, \vek k$ and
$ \Omega_k$ is the precessional frequency of $\vek k$.
It should be obvious that 
$\dot {\vek e}_{3} = \vek \Omega_e \times \vek e_{3} $  is identical to
$\dot {\vek e}_{3} =  \vek \Omega_k \times \vek e_{3}$ 
as $ \vek e_{3} \equiv \vek k$.
It is possible to
construct a $\vek k$-based 
 co-moving triad ($\vek n, \vek \xi=\vek k \times \vek n, \vek k$) and introduce an
orbital phase $\Phi$ such that 
\begin{subequations}
\label{Eq_n_xi}
\begin{align}
 \vek n  &= \cos \Phi \, \vek e_1 + \sin \Phi \, \vek e_2\,, \\  
\vek \xi &= - \sin \Phi \, \vek e_1 + \cos \Phi \, \vek e_2 \,.
\end{align}
\end{subequations}
It is fairly straightforward to obtain following 
expressions for 
the time derivatives of $\vek n$ and $\vek \xi$ 
\begin{subequations}
\begin{align}
\label{Eq_ndot_xidot}
\dot { \vek n}  &= \dot {\Phi}\, \vek \xi + \vek \Omega_e \times \vek n \,,\\ 
\dot {\vek \xi} &= - \dot {\Phi}\, \vek n + \vek \Omega_e \times \vek \xi \,.
\end{align}
\end{subequations}
 
  We are now in a position to obtain the differential equation for $\Phi$.
This is derived with the help of the frame independent adiabatic condition for circular orbits, namely
$ \dot { \vek n} \cdot \dot { \vek n} \equiv \omega^2$.
Employing the above expression for $ \dot { \vek n}$ in such an adiabatic condition leads
to 
\begin{equation}
\label{Eq_omega2}
  \omega^2 = \dot{\Phi}^2 + \Omega_{e\xi}^2 \,,
\end{equation}
where $\Omega_{e\xi} =\vek \Omega_e \cdot \vek \xi  $ is given by
\begin{equation}
\label{Eq_Omega_e_xi}
\Omega_{e\xi}
            =\frac{c^3}{Gm}\, x^3\,\Bigg\{ \delta_1\,q\, \chi_1\, 
\left( \vek s_1 \cdot \vek \xi \right) +\frac{\delta_2}{q}\, \chi_2\, \left( \vek s_2 \cdot \vek \xi \right) \Bigg\} \,.
\end{equation}
This results  in the following 3PN accurate differential equation for $\Phi$
\begin{equation}
\label{Eq_phidot}
 \dot {\Phi} = \frac{c^3}{G\, m}\, x^{3/2} \, \biggl\{  1- \frac{x^3}{2}\, \Bigl[\delta_1 \, q\, \chi_1 \, (\vek s_1\cdot \vek \xi) 
 + \frac{\delta_2}{q}\, \chi_2\, (\vek s_2\cdot \vek \xi)\Bigr]^2\biggr\} \,.
\end{equation}
These additional terms appear at the 3PN order as we employ equation~(\ref{Eq_l_omega_k}) for $\vek \Omega_k$
that only 
incorporates the leading order spin-orbit interactions appearing at 1.5PN order.
It is not difficult to deduce that the inclusion of 2PN order spin-spin interaction terms in $\vek \Omega_k$
can lead to certain 4PN order contributions to the $\dot {\Phi}$ equation.
Additionally, the presence of the above 3PN 
contributions to $\dot {\Phi}$ equation 
may be  attributed to the fact that the orbital velocity $\vek v$ can have
 non-vanishing PN order components along $\vek L$ 
while considering generic spinning compact binaries \cite{GG1}.

   The use of $\vek L$ to describe binary orbits also modifies the evolution 
 equation for $\omega$ (or $x$). This is because  
the spin-orbit interactions are usually incorporated in terms of $\vek s_1 \cdot \vek l$ and $\vek s_2 \cdot \vek l$,
in the literature \cite{LK_95,BCV}.
These terms require modifications due to the following 1.5PN order relation connecting $\vek l$ and $\vek k$ 
%
\begin{eqnarray}
\label{Eq_l_k}
 \vek l &=& \vek k + x^{3/2}\, \biggl \{-\frac{1}{2}\, \eta \, \bigl [\chi_1\, (\vek s_1 \cdot \vek n)+ \chi_2\, (\vek s_2 \cdot \vek n) \bigr]\, \vek n    \nonumber \\
          &&+ \bigl [2\, X_1^2 \, \chi_1 \, (\vek s_1 \cdot \vek \xi)+ \eta\, \chi_1\, (\vek s_1 \cdot \vek \xi)  \nonumber \\
          &&+ 2\, X_2^2 \, \chi_2 \, (\vek s_2 \cdot \vek \xi)+ \eta\, \chi_2\, (\vek s_2 \cdot \vek \xi) \bigr]\, \vek \xi \biggr\} \,,
\end{eqnarray}
where $X_1=m_1/m$ and $X_2=m_2/m$.
The above relation can easily be extracted, for example, from  equations~(6.10) and (7.10) in \cite{BBF}.
 This relation leads to certain additional 3PN order contributions to $\dot{x}$ 
while describing the binary orbits with $\vek k$.
These additional terms are, for example, with respect to equation (3.16) of \cite{Bohe2013}
that provides PN accurate expression for $dx/dt$ while invoking $\vek l$ to describe
binary orbits.  Our additional contributions to $\dot x$ appear at 3PN order
as the dominant spin-orbit interactions, 
in terms of 
$\vek s_1 \cdot \vek l$ and $\vek s_2 \cdot \vek l$, 
contribute to
the $x$ evolution equation at 1.5PN order.
The differential equation for $x$ in our approach may be written as  
\begin{align}
\label{Eq_dxdt}
\frac{ d x}{dt} &= \frac{dx}{dt}(Equation~(3.16)\, in \, \cite{Bohe2013}; \, \vek l \rightarrow \vek k) \nonumber \\
&\quad+\frac{64}{5}\frac{c^3}{Gm}\eta\, {x}^5 
\biggl \{
x^3\, \biggl[
 -\frac{47}{3}\, X_1^2\, \chi_1\, \Bigl[-\frac{1}{2}\, \eta\, \Bigl(\chi_1\, (\vek s_1 \cdot \vek n)+\chi_2\, (\vek s_2 \cdot \vek n)\Bigr)(\vek s_1 \cdot \vek n) \nonumber \\
 &\quad+\Bigl(2\, X_1^2\, \chi_1\, (\vek s_1\cdot \vek \xi)+\eta\, \chi_1\, (\vek s_1 \cdot \vek \xi)  
 +2\, X_2^2\, \chi_2\, (\vek s_2\cdot \vek \xi) 
 +\eta\, \chi_2\, (\vek s_2 \cdot \vek \xi) \Bigr) (\vek s_1 \cdot \vek \xi)\Bigr]  \nonumber \\
 &\quad-\frac{47}{3}\, X_2^2\, \chi_2\, \Bigl[-\frac{1}{2}\, \eta\, \Bigl(\chi_1\, (\vek s_1 \cdot \vek n)  
 +\chi_2\, (\vek s_2 \cdot \vek n)\Bigr)(\vek s_2 \cdot \vek n)   \nonumber \\
 &\quad+\Bigl(2\, X_1^2\, \chi_1\, (\vek s_1\cdot \vek \xi) 
 +\eta\, \chi_1\, (\vek s_1 \cdot \vek \xi) 
 +2\, X_2^2\, \chi_2\, (\vek s_2\cdot \vek \xi)  
 +\eta\, \chi_2\, (\vek s_2 \cdot \vek \xi) \Bigr) (\vek s_2 \cdot \vek \xi)\Bigr]  \nonumber \\
 &\quad-\frac{25}{4}\, (X_1-X_2)\, X_2\, \chi_2\, \Bigl[-\frac{1}{2}\, \eta\, \Bigl(\chi_1\, (\vek s_1 \cdot \vek n) 
 +\chi_2\, (\vek s_2 \cdot \vek n)\Bigr)(\vek s_2 \cdot \vek n)   \nonumber \\
 &\quad+\Bigl(2\, X_1^2\, \chi_1\, (\vek s_1\cdot \vek \xi)+\eta\, \chi_1\, (\vek s_1 \cdot \vek \xi) 
 +2\, X_2^2\, \chi_2\, (\vek s_2\cdot \vek \xi)
 +\eta\, \chi_2\, (\vek s_2 \cdot \vek \xi) \Bigr) (\vek s_2 \cdot \vek \xi)\Bigr]  \nonumber \\
 &\quad+\frac{25}{4}\, (X_1-X_2)\, X_1\, \chi_1\, \Bigl[-\frac{1}{2}\, \eta\, \Bigl(\chi_1\, (\vek s_1 \cdot \vek n)+\chi_2\, (\vek s_2 \cdot \vek n)\Bigr)(\vek s_1 \cdot \vek n)  \nonumber \\
&\quad+\Bigl(2\, X_1^2\, \chi_1\, (\vek s_1\cdot \vek \xi)+\eta\, \chi_1\, (\vek s_1 \cdot \vek \xi) 
 +2\, X_2^2\, \chi_2\, (\vek s_2\cdot \vek \xi) 
 +\eta\, \chi_2\, (\vek s_2 \cdot \vek \xi) \Bigr) (\vek s_1 \cdot \vek \xi)\Bigr] \biggr] 
\biggr \} \,,
\end{align}
where the first term is adapted from equation~(3.16) in \cite{Bohe2013} by replacing 
its $\vek l$ vectors by our $\vek k$ vectors.
Clearly, these additional {\it spin}-squared terms contribute to $dx/dt$ at 3PN order.
Note that the use of $\vek k$ 
for $\vek l$ in 
the leading order spin-spin contributions to $dx/dt$, as listed in equation~(1) of \cite{BCV},
results in additional 3.5PN order  {\it spin}-cubed terms.
Such contributions to our $dx/dt$ equation are neglected as
the {\scshape lalsuite} SpinTaylorT4 code that
implement the $\vek L_{\rm N}$-based precessing convention does not
include any {\it spin}-cubed terms in its differential equation for $x$.

  In what follows, we model inspiral GWs from spinning compact binaries in our 
$\vek k$-based precessing convention. 
We begin by displaying the following quadrupolar order expressions for the two GW polarization states:
\begin{subequations}
\label{Eq_hp_hx}
\begin{eqnarray}
 h_{\times}|_{\rm Q}(t) &=&  2\, \frac{ G\, m\, \eta \, x}{c^2\, R'}\, (2\, \xi_x\, \xi_y -2\, n_x\, n_y)\,, \\
 h_{+}|_{\rm Q}(t) &=&   2\, \frac{ G\, m\, \eta \, x}{c^2\, R'}\, (\xi_x^2 - \xi_y^2- n_x^2 + n_y^2)     \,,
\end{eqnarray}
\end{subequations}
where $\xi_{x,y}$ and $n_{x, y}$ are the $x$ and $y$ components of $\vek \xi$ and $\vek n$ in an inertial frame
associated with  $\vek N$,
the unit vector that points from the source to the detector,  while $R'$ is the  distance to the binary.
These  $x$ and $y$ components can be expressed in terms of the  Cartesian 
components of $\vek e_1$ and $\vek e_2$ via equations~(\ref{Eq_n_xi}). 
We note that the above expressions for the quadrupolar order $h_{\times}$ and $h_+$  are written in the so-called 
frame-less convention \cite{LK_95}.
These expressions emerge from the following standard definitions for the quadrupolar order GW polarization states:
\begin{subequations}
\begin{eqnarray}
 h_{\times}|_{\rm Q}(t) &=& \frac{1}{2}\,(p^i\, q^j + q^i\, p^j)\, h^{\rm TT}_{ij}|_{\rm Q}\,, \\
 h_{+}|_{\rm Q}(t) &=&  \frac{1}{2}\,(p^i\, p^j - q^i\, q^j)\, h^{\rm TT}_{ij}|_{\rm Q}\,,
\end{eqnarray}
\end{subequations}
where  $\vek p$ and $\vek q$ are the two polarization vectors forming, along
with $\vek N$,
an orthonormal right-handed triad. 
To obtain equation~(\ref{Eq_hp_hx}), we used the following
 expression for the quadrupolar order 
transverse-traceless part of the far-zone field $h^{\rm TT}_{ij}|_{\rm Q}$ 
\begin{align}
 h^{\rm TT}_{ij}|_{\rm Q}= \frac{4\, G\, m\, \eta\, x}{c^2\, R'}\, (\xi^i\, \xi^j- n^i\, n^j)\,.
\end{align}
In the frame-less convention, we let the components of $\vek p$ and $\vek q$ in the $\vek N$-based inertial frame
to be $\vek p=(1,0,0)$ and $\vek q=(0,1,0)$. 
Clearly, we need to specify how the Cartesian components of $\vek \xi$ and $\vek n$ vary in time 
to obtain temporally evolving GW polarization states
for inspiralling generic  
spinning compact binaries. 
Therefore, we require to solve numerically the differential equations for 
$\Phi, x, \vek e_1$ and $\vek e_2$
to obtain $h_{\times}|_{\rm Q}(t)$ and $h_{+}|_{\rm Q}(t)$.
We use equation~(\ref{Eq_phidot}) for $\Phi$ while the differential equation for $x$ is given by equation~(\ref{Eq_dxdt})
that contains all the non-spinning contributions accurate up to 3.5PN order and the usual spin contributions that are fully 2PN accurate.
These contributions are provided, for example, by equation~(3.16) in \cite{Bohe2013} and are listed in 
the \ref{appendix}, where 
we have replaced $\vek l$ by $\vek k$.
These specific PN order choices are influenced by the fact that 
the {\scshape lalsuite} SpinTaylorT4 code
also employs 
an evolution equation for $\omega$ that incorporates such PN contributions.
We note that this routine implements the $\vek l$-based {precessing convention} of \cite{BCV}
to construct inspiral templates to search for GWs from generic spinning binaries.
The differential equations for $\vek e_1$ and $\vek e_2$ in our approach are given by
\begin{subequations}
\label{Eq_e1_e2_dot}
\begin{eqnarray}
\label{Eq_e1dot}
 \dot{\vek e_1} &=& \vek \Omega_e \times \vek e_1 = (\vek \Omega_k - (\vek \Omega_k \cdot \vek k))\times \vek e_1\,, \\
 \dot{\vek e_2} &=& \vek \Omega_e \times \vek e_2 = (\vek \Omega_k - (\vek \Omega_k \cdot \vek k))\times \vek e_2\,,
\end{eqnarray}
\end{subequations}
where we use the following expression for $\vek \Omega_k$ that incorporates the leading order spin-orbit
and spin-spin interactions
\begin{align}
\label{Eq_Omega_k}
 \vek \Omega_k &=\frac{c^3}{Gm}\, x^{3}\,\Bigg\{ \delta_1\,q\, \chi_1\,  \vek s_1  
+\frac{\delta_2}{q}\, \chi_2\, \vek s_2   
-\frac{3}{2}\,x^{1/2}\,\eta\, \chi_1\, \chi_2\, \bigl[ (\vek k \cdot \vek s_1)\, \vek s_2 
+ (\vek k \cdot \vek s_2 )\, \vek s_1    \bigr] \Bigg\}  \,.
\end{align}
It is fairly straightforward to verify that this expression is identical to 
the coefficient of $\hat { \bf L}_{\rm N}$ appearing on the right hand side of equation~(9) in \cite{BCV}.
The above equations imply that we also need to invoke 
the precessional equations for $\vek s_1$, $\vek s_2$ and $\vek k$ (or $\vek e_3$) to tackle numerically the dynamics of such binaries.
The three coupled equations  for
$\vek s_1$, $\vek s_2$ and $\vek k$ that include the leading order spin-orbit
and spin-spin interactions read
\begin{subequations}
\label{eq:s1_s2_dot}
\begin{align}
\label{eq:kdot}
{\dot {\vek s}_{1}} &=  \frac{c^3}{Gm}\, x^{5/2}\,
\Bigg\{\delta_1 \left(\vek k\times \vek s_1\right)  
+ \frac{1}{2}\,x^{1/2}\, \bigg[ X_2^2\,\chi_2\,(\vek s_2 \times \vek s_1)    
-3\, X_2^2\,\chi_2\,(\vek k \cdot \vek s_2) \, (\vek k \times \vek s_1) \bigg] \Bigg\} \,, \\
{\dot {\vek s}_{2}} &=  \frac{c^3}{Gm}\,x^{5/2}\,
\Bigg\{\delta_2 \left(\vek k\times \vek s_2\right)  
+ \frac{1}{2}\,x^{1/2}\,\bigg[  X_1^2\,\chi_1\,(\vek s_1 \times \vek s_2)    
-3\,  X_1^2\,\chi_1\,(\vek k \cdot \vek s_1) \, (\vek k \times \vek s_2) \bigg] \Bigg\} \,, \\
{\dot {\vek k}} &=  \frac{c^3}{Gm}\, x^{3}\,\Bigg\{ \delta_1\,q\, \chi_1\, 
\left( \vek s_1\times\vek k\right)  
+\frac{\delta_2}{q}\, \chi_2\, \left( \vek s_2\times\vek k\right)   \nonumber \\
&\quad-\frac{3}{2}\,x^{1/2}\,\eta\, \chi_1\, \chi_2\, \biggl[ (\vek k \cdot \vek s_1)\, (\vek s_2 \times \vek k) 
+ (\vek k \cdot \vek s_2 )\, (\vek s_1 \times \vek k)   \biggr] \Bigg\} \,.
\end{align}
\end{subequations}
It is not very difficult to verify that the above equations for 
${\dot {\vek s}_{1}}$ and ${\dot {\vek s}_{2}}$ are 
identical to equations~(2) and (3) in \cite{BCV} while 
 the differential equation for $\vek k$, as expected, arises from the usual conservation of total angular momentum $\vek J$.
This conservation implies that  $L\,\dot{\vek k}=-S_1\, \dot{\vek s_1}-S_2\, \dot{\vek s_2}$.
We would like to state again that the equations (\ref{Eq_phidot}), (\ref{Eq_dxdt_l}), (\ref{eq_sk}) and (\ref{Eq_dxdt})
provide the differential equations for $\Phi$ and $x$ in the present implementation of $\vek k$-based
precessing convention. Strictly speaking, the use of equation (\ref{Eq_Omega_k}) for $\vek \Omega_k$ requires us to 
include the additional 4PN and 3.5PN  contribution to $d\Phi/dt$ and $dx/dt$, respectively. 
However, we do not incorporate such spin-quartic and spin-cubic terms in our present work.

  In practice, we 
numerically solve simultaneously  
 the differential equations for $\vek e_1, \vek k$, $\vek s_1$, $\vek s_2$, $\Phi$ and $x$ to
obtain temporally evolving Cartesian components of $\vek n$ and $\vek \xi$.
The resulting variations in these Cartesian components 
are imposed on the  expressions
 for $h_{\times,+}|_{\rm Q}(t)$, given by equations~(\ref{Eq_hp_hx}). This leads to  inspiral waveforms
for generic spinning compact binaries in our $\vek k$-based precessing convention.
Note that we 
 do not solve the differential equation for $\vek e_2$. This is because the temporal evolution of 
 $\vek e_2$ can be estimated using the relation $\vek e_2(t)=\vek k(t)\times \vek e_1(t)$.
This implies that we solve 12 differential equations for the Cartesian components 
of $\vek e_1$, $\vek k$, $\vek s_1$ and $\vek s_2$
 along with the differential equations for $\Phi$ and $x$ to track the time evolution for 
 the Cartesian components of $\vek n$ and $\vek \xi$.
The required initial values for the Cartesian components of $\vek e_1$, $\vek k$, $\vek s_1$ and $\vek s_2$ are given by freely 
choosing the following five angles: $\theta_{10}$, $\phi_{10}$, $\theta_{20}$, $\phi_{20}$ and $\iota_0$.
These angles specify the above four unit vectors in the $\vek N$-based inertial frame at the initial epoch such that 
\begin{subequations}
\label{eq:16}
\begin{align}
\label{eq:s1_s2_1}
\vek s_1 &= \left ( \sin \theta_{10}\,\cos \phi_{10}, \sin \theta_{10}\,\sin \phi_{10}, \cos \theta_{10} \right )\,,\\  
\label{eq:s1_s2_2}
\vek s_2 &= \left ( \sin \theta_{20}\,\cos \phi_{20}, \sin \theta_{20}\,\sin \phi_{20}, \cos \theta_{20} \right )\,,
\\
\vek k &= (\sin \iota_0 ,0 ,\cos \iota_0)\,,\\
\vek e_1 &= (\cos \iota_0 ,0 ,-\sin \iota_0)\,.
\end{align}
\end{subequations}
This choice is also influenced by the {\scshape lalsuite} SpinTaylorT4 code of LSC.
Additionally, we let the initial $x$ value to be  $x_0=(G\, m\, \omega_0/c^3)^{2/3}$ where $\omega_0=10\pi$ Hz
and the initial phase $\Phi_0$ to be zero.

 We move on to explain how to specify initial conditions, physically identical to equations~(\ref{eq:16}), 
 while constructing inspiral waveforms based on the $\vek l$-based precessing convention.
Clearly, the initial orientations of two spin vectors in the inertial $\vek N$  frame should be identical
in the two approaches. 
However, the orientation of $\vek l$ in such an inertial frame is different from that of $\vek k$.
We compute $\vek N \cdot \vek l$ using equation~(\ref{Eq_l_k}) that provide 1.5PN accurate relation connecting 
$\vek l$ and $\vek k$.
This PN relation makes the value of $\vek N \cdot \vek l$ at $x_0$ to depend on 
$\iota_0, m, \eta, \chi_1, \chi_2, \Phi_0$ and the four angles that specify $\vek s_1$ and $\vek s_2$ 
in  the inertial $\vek N$  frame. We observe that 
[$\vek N \cdot \vek l (x_0) - \vek N \cdot \vek k (x_0)$] is maximum for equal mass maximally spinning 
compact binaries and the difference is usually less than $0.1 \%$.
The value of $\vek N \cdot \vek l (x_0) $ specifies the initial orientation of $\vek l$ in the inertial $\vek N$  frame
as we usually let its azimuthal angle to be zero along with $\Phi_p(x_0)$
(we have verified that the changes in these angles play no role in our match computations).
The difference in $\vek N \cdot \vek l (x_0)$ and $\vek N \cdot \vek k (x_0)$ values leads to slightly different 
values for $\vek k \cdot \vek s_1$ and $\vek l \cdot \vek s_1$ at the initial epoch.
Therefore, the orbital frequency and phase evolutions are slightly different in the two approaches even in the absence of 
our 3PN order additional terms. 
We observe that the differences in these two dot products, namely  $\vek k \cdot \vek s_1$ and $\vek l \cdot \vek s_1$,
have their maximum values for equal mass maximally spinning compact binaries.
In the next section,  we purse the {\it match} computations to 
probe the implications of additional 3PN order terms present in the 
frequency and phase evolution equations while 
constructing our $\vek k$-based  inspiral waveforms.

\section{ Match computations involving the above two families of inspiral waveforms}
\label{result}
 We employ the {\it match}, detailed in \cite{DIS98, BO96}, to compare inspiral waveforms constructed via
the above described $\vek l$ and $\vek k$-based precessing conventions.
Our comparison is influenced (and justified) by the fact that the precessing source frames of the above 
two conventions are functionally identical.
This should be evident from the use of {\it the same} precessional frequency, appropriate
for $\vek k$, to obtain PN accurate expressions for both the $\vek l$-based $\vek \Omega_e^l $ and 
$\vek k$-based $\vek \Omega_e $.
Therefore, the {\it match} estimates probe influences of the additional 3PN order
terms present in the differential equations for $\Phi$ and $x$ in our approach.
Additionally, we have verified that these 3PN order terms are not present in 
the usual implementation of the precessing
convention as provided by the {\scshape lalsuite} SpinTaylorT4 code.

 Our match ${\cal M}(h_l, h_k)$ computations involve $h_l $ and $ h_k$, the two families of inspiral
waveforms arising from the $\vek l$ and $\vek k$-based precessing conventions.
The $ h_l$ inspiral waveform families are adapted from the {\scshape lalsuite} SpinTaylorT4 code of LSC
while $ h_k$ families, as expected, arise from our approach (equations~(\ref{Eq_hp_hx})).
It should be noted that we employ  the quadrupolar (Newtonian)  order expressions
for $h_{\times, +}$ while computing $h_l$ and  $h_k$ in the present analysis.
However, the {\scshape lalsuite} SpinTaylorT4 code can provide waveforms that include 1.5PN order corrections to their amplitudes.
We would like to stress that the two families involved in our ${\cal M}(h_l, h_k)$ computations are characterized by 
identical values of $m, \eta, \chi_1 $ and $ \chi_2$. Additionally, the initial
orientations of the two spins in the $\vek N$-based inertial frame were also chosen to be identical.
The computation of $\vek N \cdot \vek l$ from $\vek N \cdot \vek k$ with the help of  equation~(\ref{Eq_l_k})
ensures that $\vek l$ and $\vek k$  orientations  at the initial epoch are physically equivalent.
Therefore, our  ${\cal M}(h_l, h_k)$ computations indeed compare two waveform families with physically 
equivalent orbital and spin configurations at the initial epoch.
To obtain a specific ${\cal M}(h_l, h_k)$ estimate,  we first compute an overlap function between
the relevant $h_l $ and $ h_k$ inspiral waveforms: 
\begin{equation}
 \mathcal{O}(h_l, h_k) =   < \hat h_l, \hat h_k>  =
\frac{\langle h_l|h_k \rangle}{\sqrt{\langle h_l|h_l \rangle \, \langle h_k|h_k\rangle}} \,,
\end{equation}
where $\hat h_l$ and $\hat h_k$ stand for the normalized $h_l(t)$ and $h_k(t)$ waveforms, respectively.
The angular bracket between $h_l$ and $h_k$ defines certain  noise 
weighted inner product, namely
\begin{equation}
\langle  h_l |  h_k \rangle= 4\, {\rm Re}\,  \int_{f_{\rm low}}^{f_{\rm cut}} \, 
\frac{\tilde h_l^*(f)\, \tilde h_k(f)}{S_{\rm h}(f)} df \,.
\end{equation}
In the above equation, $\tilde h_l(f)$ and $\tilde h_k(f)$ stand for 
the Fourier transforms of $h_l(t)$ and
$h_k(t)$, while  
$S_{\rm h}(f)$ denotes 
the one-sided power spectral density. We have invoked the 
zero-detuned, high power sensitivity curve of aLIGO \cite{LIGO_2010} in our ${\cal M}(h_l, h_k)$ computations.
The upper cut-off
frequency $f_{\rm cut}$ is chosen to be $c^3/(G\, m\, \pi\, 10^{3/2})$ while the lower cut-off
frequency $f_{\rm low}$, associated with the GW detector, equals $10$ Hz.
The match ${\cal M}(h_l,h_k)$ is computed by maximizing the $\mathcal{ O}(h_l, h_k)$ over 
two extrinsic variables, namely
the time of arrival $t_0$ and the 
overall phase $\phi_0$ of GW at time $t_0$ \cite{Veitch14}.
This leads to
\begin{equation}
\label{Eq_match}
{\cal M }= \max_{t_0, \phi_0}\, \mathcal{O}(h_l, h_k)\,.
\end{equation}
The maximizations of  over $t_0$ and $\phi_0$ are performed by 
following \cite{DIS98}. We perform the maximization over $t_0$ via the FFT algorithm 
while the maximization over $\phi_0$ requires us to deploy two orthogonal templates.
Let us emphasize that we 
terminate  $h_l $ and $ h_k$ inspiral waveform families when their respective
$x$ parameters reach $0.1$ which roughly corresponds to orbital separations $\sim 10 \, G\, m/c^2 $.
This choice  arguably ensures the validity of PN approximation to describe the temporal 
evolutions of the above two families 
in our $[f_{\rm low}$-$f_{\rm cut}]$ frequency window.
Therefore, it is reasonable to associate the departure of 
${\cal M}$ estimates from unity to 
the additional 3PN order contributions to the differential equations for the orbital phase and 
the associated angular frequency.

\begin{figure}[!ht]
\begin{center}
$\begin{array}{ccc}
 \includegraphics[width=62mm,height=60mm]{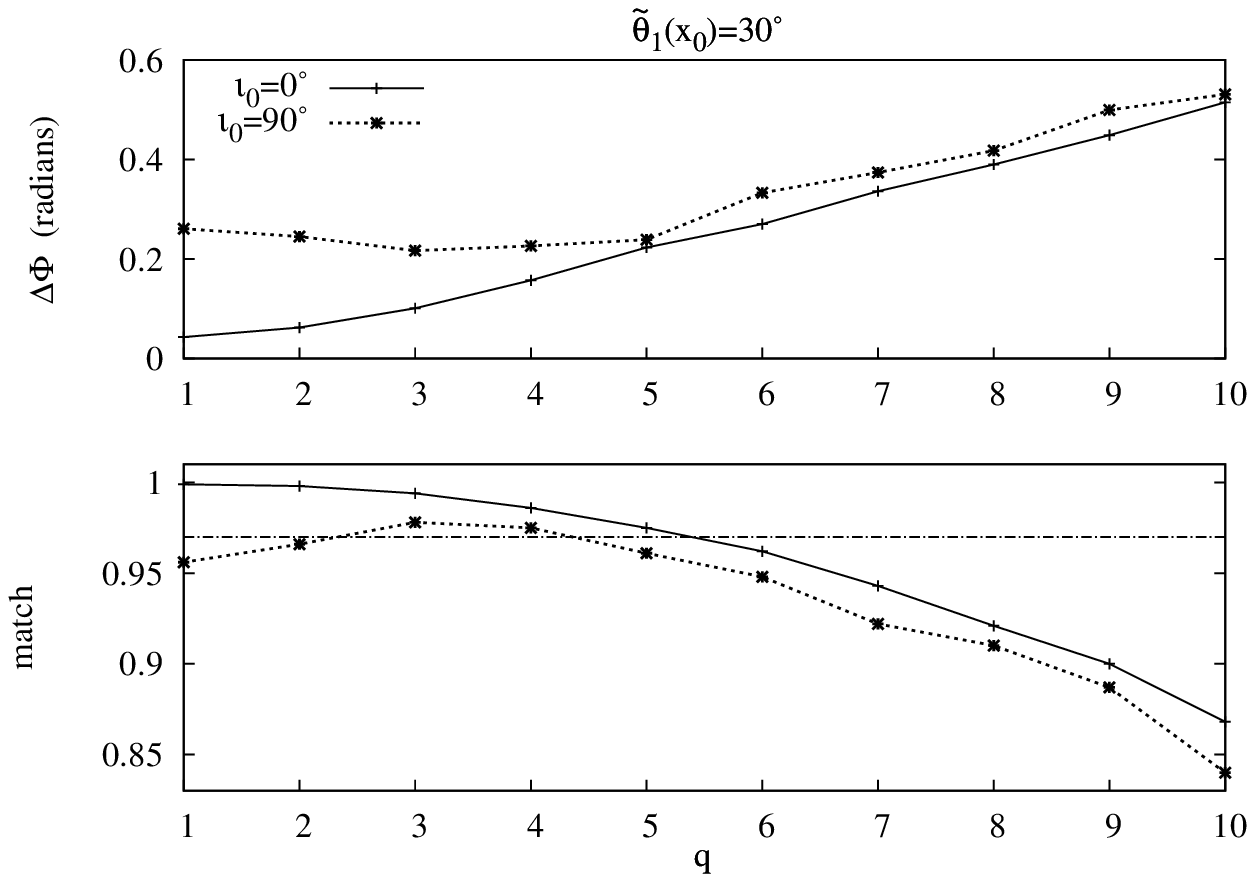}&
 \includegraphics[width=62mm,height=60mm]{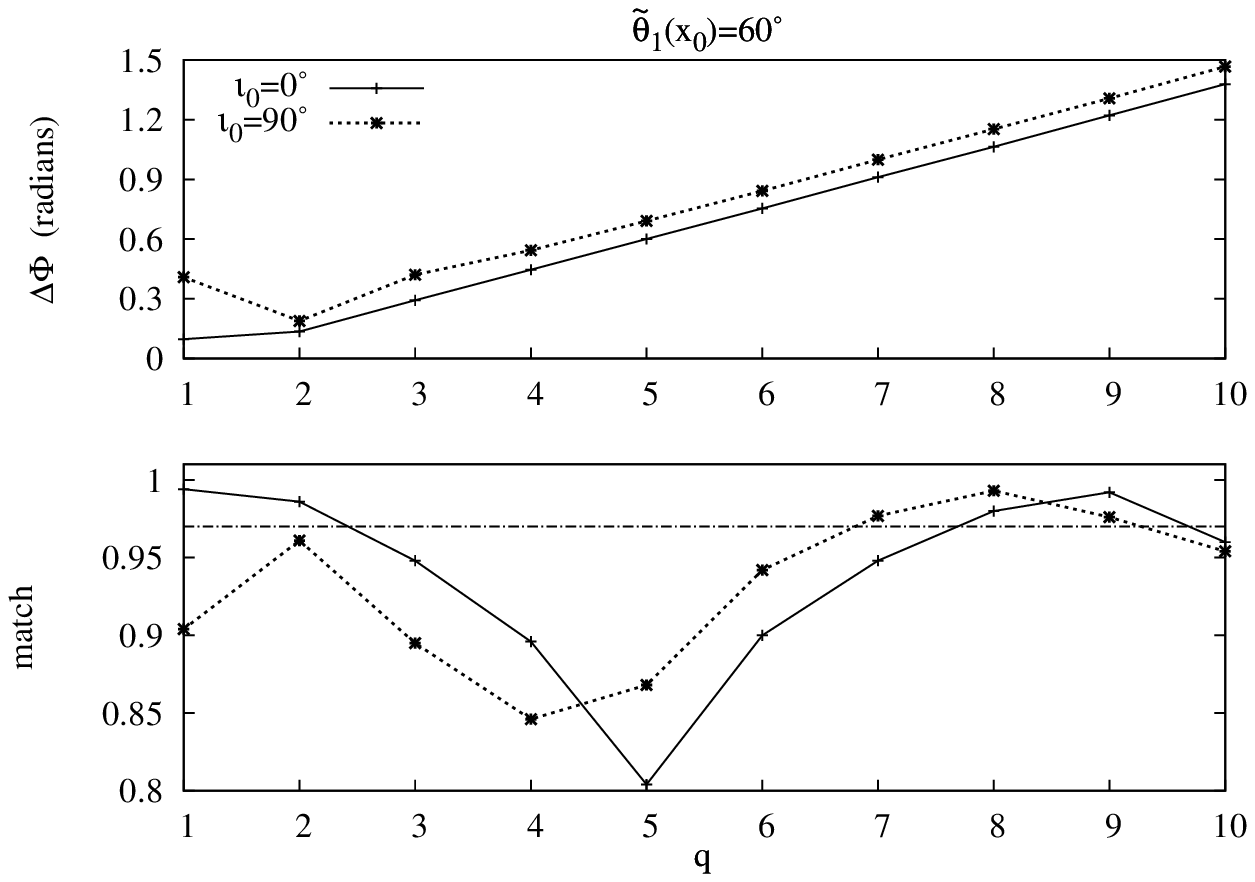}
\end{array}$
\end{center}
\caption{Plots for the  accumulated orbital phase ($\Delta
\Phi$) and the associated match ($\mathcal{M}$) estimates
as functions of the mass ratio $q$ for maximally spinning  $m=30M_{\odot}$
BH binaries inspiralling in the $[f_{\rm low}$-$f_{\rm cut}]$ frequency interval.
The left and right panel plots are for binaries having dominant 
spin-orbit misalignments $\tilde\theta_1=30^{\circ}$ and $60^{\circ}$, respectively, at $x_0$.
The solid line plots in each panel correspond to face-on binaries
whereas the dotted line curves are for edge-on binaries.
We find a gradual degradation of the $\mathcal{M}$ estimates for
higher $q$ values
for $\tilde\theta_1=30^{\circ}$ binaries.
However, this trend is not visible for $\tilde\theta_1=60^{\circ}$
binaries.
These configurations with different spin-orbit misalignments and binary orientations 
are obtained by setting different values for $\theta_{10}$ and $\iota_0$. Face-on binaries with 
$\tilde\theta_1(x_0)=30^{\circ}$ and $60^{\circ}$ have 
$(\theta_{10}, \iota_0)$ as $(30^{\circ}, 0^{\circ})$ and $(60^{\circ}, 0^{\circ})$
while the edge-on binaries have 
$(\theta_{10}, \iota_0)$ as $(60^{\circ}, 90^{\circ})$ and $(30^{\circ}, 90^{\circ})$, respectively.
The initial values of the other angular parameters remain the same, namely
$\phi_{10}=0^{\circ}$,
$\theta_{20}=20^{\circ}$, $\phi_{20}=90^{\circ}$.
The PN accurate relation between $\vek k$ and $\vek l$ also results in 
slightly different dominant spin-orbit 
misalignments in the two approaches.
}
\label{figure:q_M_Phi}
\end{figure}

  We move on to list results of some of our ${\cal M}$ computations.
It is not very difficult to realize that extensive ${\cal M}$ computations that deal
with all aLIGO relevant spin and binary configurations will be rather difficult to achieve.
Therefore, we restrict our attention to a selected number of binaries to compare
the  $h_l(t)$ and $h_k(t)$
inspiral waveform families.
 In our match computations, we mainly consider binaries having total mass $m\geq30M_{\odot}$
due to the following two reasons. First, it is comparatively  expensive (computationally) 
to generate lengthy inspiral 
waveforms for low mass binaries in the aLIGO frequency window.
Additionally, we are interested in to explore the dependence of our ${\cal M}$
estimates on the mass ratio $q$  in the $[1-10]$ range.
Clearly, low $m$ binaries can lead to secondary BHs having masses lower than the usual neutron star masses
for high $q$ cases.
In figure~\ref{figure:q_M_Phi}, we consider 
maximally spinning BH binaries, characterized by the total mass $m=30M_{\odot}$,
while varying the mass ratio $q$ from unity to $10$.
The left and right panel plots are for configurations having initial dominant spin-orbit misalignments 
$\tilde\theta_1(x_0)$ $(\cos^{-1}(\vek k \cdot \vek s_1))$ given by $30^{\circ}$ and $ 60^{\circ}$, respectively.
Additionally,  we let the initial orbital plane orientation in the $\vek N$-based inertial frame to take 
two values leading to edge-on ($\iota_0= 90^{\circ}$) and face-on ($\iota_0= 0^{\circ}$) binary orientations.
These binary configurations, characterized by two different spin-orbit misalignments and orbital plane orientations 
in the $\vek N$-based inertial frame, 
are obtained by choosing appropriately different values for 
the more massive BH initial spin orientation, namely
$\theta_{1}(x_0)$ in equation~(\ref{eq:s1_s2_1}).
For example, we choose $\theta_{10}$ to be $30^{\circ}$ and $60^{\circ}$
to get $\tilde\theta_1(x_0)$ equals  $30^{\circ}$ and $60^{\circ}$, respectively, 
for face-on binaries.
However, in the case of edge-on binaries, we require to choose
$\theta_{10}$ to be $60^{\circ}$ and $30^{\circ}$, respectively.
All other initial angular variables, appearing in equations~(\ref{eq:s1_s2_1}) and (\ref{eq:s1_s2_2}), were chosen to be 
$\phi_{10}=0^{\circ}$,
$\theta_{20}=20^{\circ}$, $\phi_{20}=90^{\circ}$ (we have verified that the  ${\cal M}$ estimates are rather insensitive to the choice of
these initial angular variables).
 Let us note that $\vek l$ orientations (from $\vek N$) for these configurations will be slightly 
different from $0^{\circ}$ or $90^{\circ}$ due to the 1.5PN accurate relation between $\vek l$ and 
 $\vek k$.

  The upper and lower row plots in  figure~\ref{figure:q_M_Phi} are for $\Delta \Phi$, the accumulated orbital phase differences in
the $[f_{\rm low}$-$f_{\rm cut}]$ frequency interval while dealing with 
the $h_l $ and $ h_k$ inspiral waveforms
and the associated ${\cal M}(h_l,h_k)$  estimates, respectively.
We find that the variations in ${\cal M}$ estimates are quite independent of the initial 
orbital plane orientations. 
However, variation in match estimates do depend on the initial dominant spin-orbit misalignment and the mass ratio $q$.
The left panel plots show gradual decrease in  ${\cal M}$ values as we increase the $q$ value
and this variation is reflected in the gradual increase of $\Delta \Phi$.
Incidentally, this pattern is also observed for configurations having somewhat smaller initial dominant spin-orbit misalignments.
However, the ${\cal M}$ estimates are close to unity for tiny $\tilde \theta_1(x_0)$ values 
and this is  expected as precessional effects 
are minimal for such binaries.
Therefore, the effect of the above discussed additional 3PN order terms are more pronounced for 
high $q$ compact binaries having {\it moderate} dominant spin-orbit misalignments.
This picture is modified for binaries having substantial dominant spin-orbit misalignments as evident from 
the ${\cal M}$ plots in the right panel of figure~\ref{figure:q_M_Phi}.
For such binaries, the ${\cal M}$ estimates dip to a minimum and recover 
as we vary $q$ from $1$ to $10$ for both edge-on and face-on orbital plane orientations. In contrast, we observe 
 a gradual increase in $\Delta \Phi (q)$. 
The monotonic increment in $\Delta \Phi (q) $ plots is essentially due to the presence of 
$X_1-X_2$ terms in the additional 3PN order contributions to $dx/dt$, given by equation~(\ref{Eq_dxdt}).
This is because $X_1-X_2 $ terms are absent for equal mass binaries which leads to smaller 
$\Delta \Phi $ estimates for smaller $q$ value binaries.
However, it is not possible
to explain the observed ${\cal M}(q)$ variations 
purely in terms the displayed $\Delta \Phi (q)$ values,
especially when precessional effects are substantial
as in the $\tilde\theta_1(x_0)=60^{\circ}$ cases.
We observe that the initial values of angles like 
$\cos^{-1}(\vek k \cdot \vek j)$ and  $\cos^{-1}(\vek N \cdot \vek j)$ also 
influence the 
precessional modulations in the waveforms, where $\vek j$ is the unit vector along $\vek J$.
Therefore, we examined how the 
initial values of these angles vary as functions of $q$.
We find monotonic increase (decrease) in  the 
initial values of $\cos^{-1}(\vek k \cdot \vek j)$ $( \cos^{-1}(\vek N \cdot \vek j) )$ when
$q$ value is varied from 1 to 10. 
Therefore, we speculate that 
the combined effect of such variations and the non-negligible $\Delta \Phi (q) $ values 
may provide a possible explanation for the dip in match estimates around q=5.
This is because initial values of the above mentioned angles do define the way
various dot products, involving $\vek n$,
 $\vek \xi$, $\vek s_1$ and 
$\vek s_2$, vary in time. It should be noted that 
these dot products 
are present in the additional 3PN order terms in the $dx/dt$ expression 
given by equation~(\ref{Eq_dxdt}).

\begin{figure}[!ht]
\begin{center}
 \includegraphics[width=82mm,height=74mm]{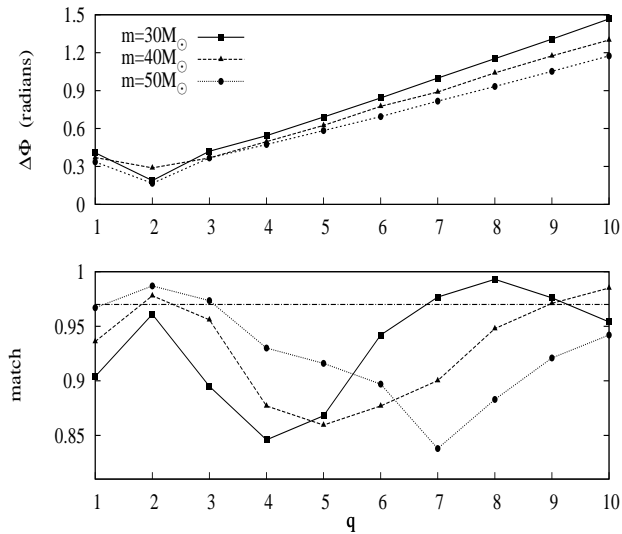}
\end{center}
\caption{
Plots that display variations in $\Delta \Phi$ and $\mathcal{M}$ as functions of $q$ 
for edge-on maximally spinning BH binaries having $m=30\,M_{\odot},40\,M_{\odot}$ and $50\,M_{\odot} $ while 
keeping $\tilde\theta_1(x_0)$ to be $60^{\circ}$.
All other initial angular parameters are identical to those used in the right panel plots of figure~\ref{figure:q_M_Phi}.
The position of the dip in $\mathcal{M}(q)$ plots is shifted towards higher $q$ values 
for higher $m$ binaries.
A possible explanation relies on the more influential contributions from the various 
dot products that appear in equation~(\ref{Eq_dxdt}) 
for such binaries having $q$ roughly in the  $4$ to $ 9$ range.
}
\label{figure:th1_M_Phi}
\end{figure}

\begin{figure}[!h]
\begin{center}
 \includegraphics[width=82mm,height=74mm]{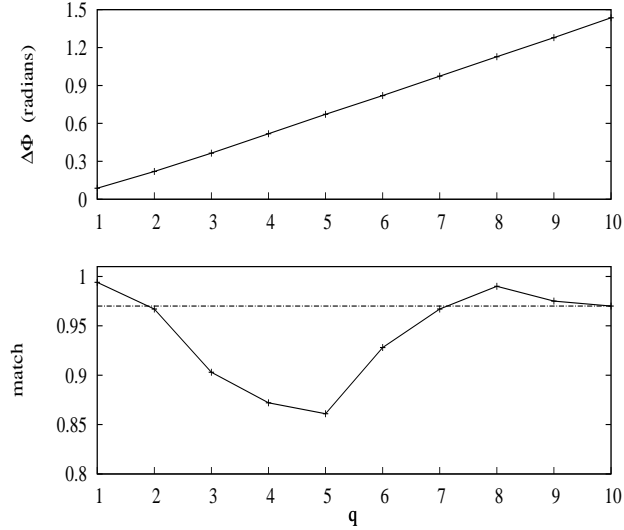}
\end{center}
\caption{
Plots that compare \emph{single-spin} SpinTaylorT4 inspiral waveform families in the $\vek l$ and 
$\vek k$-based approaches.
These $\Delta \Phi(q)$ and $\mathcal{M}(q)$ plots are for edge-on $m=30M_{\odot} $ binaries 
containing maximally spinning dominant BH having $\tilde \theta_1(x_0)=60^{\circ}$.
The qualitative and quantitative nature of these plots are essentially identical to the 
double spin binaries of  figure~\ref{figure:q_M_Phi}.
}
\label{figure:th1_M_Phi_LN}
\end{figure}

  This plausible explanation is tested in figure~\ref{figure:th1_M_Phi}, where we plot
$\Delta \Phi (q) $ and ${\cal M} (q)$ estimates for binary configurations having three different $m$ values.
It is not difficult to infer that the above listed arguments should also hold for such compact binaries.
The initial dominant spin-orbit misalignment is again chosen to be $60^{\circ}$
while considering only edge-on configurations.
All other spin and orbital orientations at the initial
epoch are identical to the cases displayed in  the right panel plots of figure~\ref{figure:q_M_Phi}.
Clearly,
the plots in figure~\ref{figure:th1_M_Phi} are qualitatively similar to those in the right panel plots of 
figure~\ref{figure:q_M_Phi}.
However, the dip in ${\cal M}$ estimates shifts to higher $q$ values 
for higher $m$ compact binaries. 
We conclude from the above two figures that 
GW data analysis relevant differences between the above two inspiral
families are more pronounced for unequal mass BH binaries.
Incidentally,  we also find similar behavior while 
plotting ${\cal M}(q)$ and $\Phi(q)$  for maximally spinning $m = 20\,M_{\odot}$ 
BH binaries. In this case, the minimum ${\cal M}$ value occurs around $q=3$ 
and this is consistent with the trend observed in  figure~\ref{figure:th1_M_Phi}.
It is reasonable to suspect that the spin-squared additional terms are influential only for maximally spinning BH binaries.
However, we observe qualitatively similar ${\cal M} (q)$ estimates for 
 binary
configurations having moderately spinning BHs ($\chi_1=\chi_2=0.75$) as well as  having mildly spinning BHs ($\chi_1=\chi_2=0.5$).  
Moreover, the ${\cal M} (q)$ computations indicate that the effect of additional 3PN order terms 
  in $d\Phi/dt$ and $dx/dt$ equations are non-negligible 
even for single-spin compact binaries. In figure~\ref{figure:th1_M_Phi_LN},  we plot 
$\Delta \Phi (q) $ and ${\cal M} (q)$ while considering $\vek l$ and $\vek k$-based single-spin waveforms 
for $m=30M_{\odot},\tilde\theta_1(x_0)=60^{\circ} $ binaries in edge-on orientations.
We find that these plots fairly resemble their double spin counterparts shown in the
right panel plots of figure~\ref{figure:q_M_Phi}.
Additionally, the variations in the initial values of $\cos^{-1}(\vek k \cdot \vek j)$ and $ \cos^{-1}(\vek N \cdot \vek j) $ 
are similar to the double spin 
binaries, while varying $q$ value from $1$ to $10$. 
Therefore, we conclude that it may be beneficial to keep the additional 3PN order
terms in the differential equations for $\Phi$ and $x$ while modeling inspiral GWs even  
from single-spin binaries
via the precessing convention.

  The LSC also developed {\scshape lalsuite} SpinTaylorT2 code that
employs a slightly different version of $dx/dt$ equation while 
implementing the $\vek l$-based precessing convention.
Note that it is possible to construct various PN approximants to model inspiral GWs
 by using the same PN accurate expressions for the 
conserved energy, far-zone energy flux and the {\it energy balance argument} \cite{Boyle_07}.
The differential equation for $x$ in the SpinTaylorT2 approximant arises from the following considerations.
We begin by displaying 
(symbolically) the 3.5PN accurate expression for $dx/dt$, employed in the {\scshape lalsuite} SpinTaylorT4
code, as 
\begin{align}
 \frac{dx}{dt}\Big |_{\rm T4}= \frac{64}{5}\frac{c^3}{Gm}\eta\, x^5\, \Big\{ 1+ A_1\, x + A_{1.5}\, x^{3/2} + A_{2}\, x^2 + A_{2.5}\, x^{5/2} + A_{3}\, x^3 + A_{3.5}\, x^{7/2} \Big\}\,,
\end{align}
where the coefficients $A_i$ are functions of $\eta, \pi, \log (6\,x) $ and $\gamma_{\rm E}$ (Euler's constant)
when including only the non-spinning contributions.
However, the coefficients from $A_{1.5}$ to $A_{3.5}$ additionally depend on $X_1, X_2, (\vek s_1 \cdot \vek l), 
(\vek s_2 \cdot \vek l),
 (\vek s_1 \cdot \vek s_2), \chi_1$ and $\chi_2$ to incorporate various spin contributions.
The explicit expressions for these $A_i$ coefficients may be obtained either from equation~(3.16) in \cite{Bohe2013} or from 
the {\scshape lalsuite} SpinTaylorT4 code itself.
With these inputs, one defines 
the differential equation for $x$ in the SpinTaylorT2 approximant to be
{\allowdisplaybreaks
\begin{align}
\label{Eq_T2}
 \frac{dx}{dt}\Big |_{\rm T2}&= \frac{64}{5}\frac{c^3}{Gm}\eta\, \frac{x^5}{ \Big\{ 1+ A_1\, x + A_{1.5}\, x^{3/2} + A_{2}\, x^2 + A_{2.5}\, x^{5/2} + A_{3}\, x^3 + A_{3.5}\, x^{7/2} \Big\}^{-1}}\,, \\
  &=\frac{64}{5}\frac{c^3}{Gm}\eta\, \frac{x^5}{ \Big\{ 1+ A'_{1}\, x + A'_{1.5}\, x^{3/2} + A'_{2}\, x^2 + A'_{2.5}\, x^{5/2} + A'_{3}\, x^3 + A'_{3.5}\, x^{7/2} \Big\}}\,.
\end{align}
}
This construction ensures that the coefficients $A'_i$ are going to depend on various $A_i$ coefficients due to the 
binomial expansion of the denominator of equation~(\ref{Eq_T2}) that includes all the  $x^{7/2}$ contributions.
For example, the $A'_{3.5}$ coefficient is going to depend explicitly on all
the $A_i$ coefficients whereas the $A'_{2}$ coefficient depends only on  the $A_1$ and $A_2$ coefficients. 
We observe that  the 
resulting spin-squared terms are usually 
 not incorporated into the $A'_3$ and  $A'_{3.5}$ coefficients in the {\scshape lalsuite} SpinTaylorT2 code of LSC.

\begin{figure}[!h]
\begin{center}
 \includegraphics[width=82mm,height=74mm]{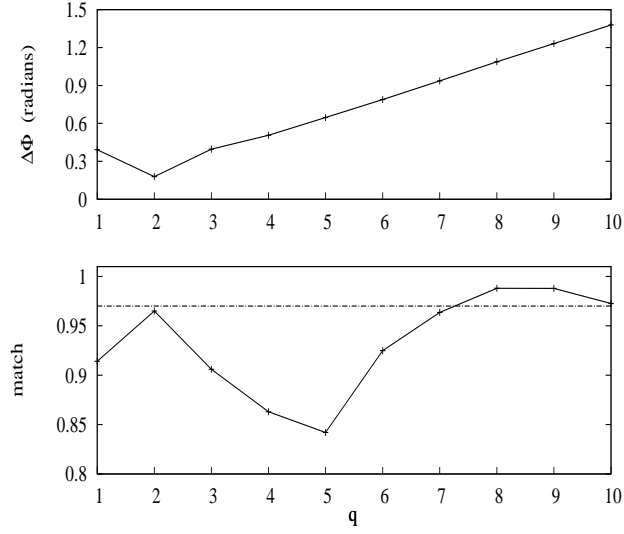}
\end{center}
\caption{ Plots for $\Delta \Phi$ and ${\cal M}$  as functions of $q$  
that invoke $\vek l$ and $\vek k$-based SpinTaylorT2 approximants.
We are dealing with edge-on maximally spinning BH binaries having $m=30\, M_{\odot} $ 
and $\tilde \theta_1 (x_0) =60^{\circ}$.
The $\Delta \Phi$ and ${\cal M}$ estimates are similar to those obtained 
using SpinTaylorT4 approximant as shown in the right panel plots of figure~\ref{figure:q_M_Phi}.
}
\label{figure:M_Phi_q_T2}
\end{figure}

  We construct the differential equation for $x$ in our SpinTaylorT2 approximant in a similar manner.
Therefore, the various $A'_i$ coefficients are given in terms of $A_i$ coefficients which can be extracted 
from our equations~(\ref{Eq_dxdt}), (\ref{Eq_dxdt_l}) and (\ref{eq_sk}) for $dx/dt$. 
For example, the expressions for $A_2$ and $A'_2$ in our approach are given by
{\allowdisplaybreaks
\begin{align}
A_2 &= \Bigl ( \frac{34103}{18144}+\frac{13661}{2016}\eta +\frac{59}{18}\eta^2 \Bigr ) 
- \frac{1}{48} \eta \chi_1 \chi_2 \,\Bigl ( 247 (\vek s_1 \cdot \vek s_2) 
-721 (\vek s_1 \cdot \vek k) (\vek s_2 \cdot \vek k)  \Bigr )  \nonumber \\
 &\quad+X_1^2\, \chi_1^2\, \Bigl( \frac{5}{2}\, (3\, ( \vek s_1  \cdot \vek k)^2-1)+\frac{1}{96}\, (7-(\vek s_1  \cdot \vek k)^2)\Bigr)  \nonumber \\
 &\quad+X_2^2\, \chi_2^2\, \Bigl( \frac{5}{2}\, (3\, (\vek s_2  \cdot \vek k)^2-1)  
  +\frac{1}{96}\, (7-(\vek s_2  \cdot \vek k)^2)\Bigr) \,, \\
A'_2&= -A_2 + A_1^2 \,, \nonumber \\
    &= -\Big [ \Bigl ( \frac{34103}{18144}+\frac{13661}{2016}\eta +\frac{59}{18}\eta^2 \Bigr ) 
- \frac{1}{48} \eta \chi_1 \chi_2 \,\Bigl ( 247 (\vek s_1 \cdot \vek s_2) 
-721 (\vek s_1 \cdot \vek k) (\vek s_2 \cdot \vek k)  \Bigr )  \nonumber \\
 &\quad+X_1^2\, \chi_1^2\, \Bigl( \frac{5}{2}\, (3\, ( \vek s_1  \cdot \vek k)^2-1)+\frac{1}{96}\, (7-(\vek s_1  \cdot \vek k)^2)\Bigr)  \nonumber \\
 &\quad+X_2^2\, \chi_2^2\, \Bigl( \frac{5}{2}\, (3\, (\vek s_2  \cdot \vek k)^2-1)  
  +\frac{1}{96}\, (7-(\vek s_2  \cdot \vek k)^2)\Bigr) \Big ]
  + \left [ -\frac{743}{336}-\frac{11\eta}{4}\right ]^2  \,. 
\end{align}
}
In the $\vek k$-based SpinTaylorT2 approximant, we incorporate our
 additional 3PN accurate spin-spin terms which contribute {\em only}
to the $A_3'$ coefficient.  
In other words, $A_3' = -A_3 + A_{1.5}^2 + 2A_1A_2 - A_1^3$ contains 
our additional 3PN order spin-squared terms via the SpinTaylorT4 based $A_3$ coefficient.
We would like to emphasize that we are not incorporating possible
3PN and 3.5PN order spin-spin terms arising due the binomial
expansion. This is to make sure that
the $\vek l$ and $\vek k$-based SpinTaylorT2 approximants differ only by our
 additional 3PN order spin-spin terms.
We explore the implications of these additional 3PN order terms again with the  help of ${\cal M}(h_l, h_k)$ computations, where 
$h_l$ and $h_k$ now stand for the  $\vek l$-based and $\vek k$-based waveform families that employ the above described 
PN accurate expressions for $dx/dt$. 
In figure~\ref{figure:M_Phi_q_T2}, we plot ${\cal M}(q)$ and $\Delta \Phi(q)$ for edge-on $m=30\,M_{\odot}$ 
maximally spinning BH binaries 
having dominant spin-orbit misalignment
equal to $60^{\circ}$ at $x_0$. 
The ${\cal M}$  and $\Delta \Phi$ estimates are similar to those obtained 
using SpinTaylorT4 approximant as shown in the right panel plots of figure~\ref{figure:q_M_Phi}.

It may be argued that the inclusion of our 3PN order spin-squared terms to
the differential equation for $x$ is not desirable.
This is because we do not incorporate 3PN order contributions to 
$dx/dt$ that arise from the next-to-leading order spin-spin interactions.
These terms are not included as they are yet to be computed in the literature.
Clearly, the present match computations should be repeated when
these 3PN order contributions are explicitly available.
 However, this should not prevent one from probing the 
data analysis implications of the adiabatic approximation
employed in the $\vek L_{\rm N}$-based precessing convention of \cite{BCV}.
This could be especially useful while invoking PN accurate differential equation for $x$ 
that incorporate higher order spin contributions as pursued in 
{\scshape lalsuite} SpinTaylorT4 and {\scshape lalsuite} SpinTaylorT2 codes, developed by
 the LSC to implement
the $\vek L_{\rm N}$-based precessing convention.
Hopefully, the data analysis implications of our 3PN order terms 
should provide additional motivation to explicitly obtain the next-to-leading
order spin-spin contributions to  $dx/dt$.
In contrast, it should be noted that the additional PN contribution to
 $d \Phi/dt$, given by equation~(\ref{Eq_phidot}), are fully accurate to 3PN order. 
The fact that such additional precessional effects  enter the expression for 
$d \Phi/dt$ only at the 3PN order should justify naming our 
approach as $\vek k$-based  precessing
convention.
It is worthwhile to remember that precession induced modulations enter
the differential equation for the orbital phase at 1.5PN order in the 
absence of precessing
convention.

\section{Conclusions}
\label{Sec_dis}

  We developed a precessing convention to model 
inspiral GWs from generic spinning compact binaries that  
employs PN accurate orbital angular momentum $\vek L$ to describe 
binary orbits and to construct the required precessing source frame.
The main motivation for our approach is   
 the usual practice of using PN accurate precessional equation, appropriate for $\vek L$, 
to evolve the Newtonian orbital angular momentum $\vek L_{\rm N}$
 while constructing inspiral waveforms.
We showed that this practice leads to
higher order PN corrections to  $\dot {\Phi}_{p} = \omega$ equation.
A set of differential equations and the quadrupolar order GW polarization states in certain
frame-less convention are developed to model inspiral GWs in our $\vek L$-based approach.
We explained why the differential equations for the orbital phase and frequency will have 
additional 3PN order terms in our approach compared to the usual 
$\vek L_{\rm N}$-based implementation of precessing convention.
The influence of these additional 3PN order terms were explored with the help of {\it match} computations
involving  $\vek L$ and $\vek L_{\rm N}$-based inspiral waveforms 
for spinning compact binaries with physically equivalent  orbital 
and spin configurations at the initial epoch.
We adapted both {\scshape lalsuite} SpinTaylorT4 and {\scshape lalsuite} SpinTaylorT2 codes, developed by
 the LSC, while implementing
$\vek L_{\rm N}$-based precessing convention.
The resulting match estimates indicate that  our additional 3PN order terms 
should not be neglected for a substantial fraction of unequal mass BH binaries. 

 It will be useful to pursue our match computations for an extended range of the relevant
parameter space. 
The present computations should also be extended by invoking  
 1.5PN order amplitude corrections to
both families of inspiral waveforms in the frame-less convention   
(the {\scshape lalsuite} SpinTaylorT4 and {\scshape lalsuite} SpinTaylorT2 codes
do incorporate such amplitude corrections).
This requires us to add certain additional 1.5PN order amplitude corrections to
the usual expressions for  $h_{\times}$ and $ h_+$ as noted in \cite{GG1}.
Investigating the influence of these additional 3PN (1.5PN) order terms in phase (amplitude)
while estimating the 
GW measurement accuracies of  
compact binary parameters will be interesting. It should also be worthwhile to probe the influence of these 
additional terms during the construction of inspiral-merger-ringdown waveforms
 from generic spinning BH binaries with the help of the effective-one-body
approach.

\ack{We thank Riccardo Sturani for helpful discussions. We are grateful to Alejandro Bohe, 
Alessandra Buonanno and Guillaume Faye for useful comments on the manuscript.  
This is a LIGO document, LIGO-P1400178.
}

\appendix

\section{3.5PN accurate expression for $dx/dt$ } 
\label{appendix}
We list below the regular contributions to $dx/dt$ that we employ while implementing our version of the SpinTaylorT4 approximant.
These contributions, adapted from \cite{Bohe2013}, read 
{\allowdisplaybreaks
\begin{align}
\label{Eq_dxdt_l}
\frac{ d x}{dt} &= \frac{64}{5}\frac{c^3}{Gm}\eta\, {x}^5 
\biggl \{
1+x \left [ -\frac{743}{336}-\frac{11\eta}{4}\right ] + x^{3/2}\,\biggl [4\, \pi -\frac{47}{3}\, s_{k} -\frac{25}{4}\, (X_1-X_2)\,\sigma_{k} \biggr ]   \nonumber \\
&\quad+x^2\,\biggl[\Bigl ( \frac{34103}{18144}+\frac{13661}{2016}\eta +\frac{59}{18}\eta^2 \Bigr ) 
- \frac{1}{48} \eta \chi_1 \chi_2 \,\Bigl ( 247 (\vek s_1 \cdot \vek s_2) 
-721 (\vek s_1 \cdot \vek k) (\vek s_2 \cdot \vek k)  \Bigr )  \nonumber \\
 &\quad+X_1^2\, \chi_1^2\, \Bigl( \frac{5}{2}\, (3\, ( \vek s_1  \cdot \vek k)^2-1)+\frac{1}{96}\, (7-(\vek s_1  \cdot \vek k)^2)\Bigr)  \nonumber \\
 &\quad+X_2^2\, \chi_2^2\, \Bigl( \frac{5}{2}\, (3\, (\vek s_2  \cdot \vek k)^2-1)  
  +\frac{1}{96}\, (7-(\vek s_2  \cdot \vek k)^2)\Bigr)\biggr]   \nonumber \\
 &\quad+ x^{5/2}\,\biggl[-\frac{4159}{672}\, \pi-\frac{5861}{144}\, s_{k}-\frac{809}{84}\, (X_1-X_2)\, \sigma_{k} \nonumber \\
 &\quad+\eta\, \Bigl( -\frac{189}{8}\, \pi +\frac{1001}{12}\, s_{k} + \frac{281}{8}\, (X_1-X_2)\, \sigma_{k}\Bigr) \biggr] \nonumber \\
&\quad + x^3\, \biggl[ \frac{16447322263}{139708800} +\frac{16}{3}\, \pi^2 -\frac{1712}{105}\, \gamma_{\rm E}-\frac{856}{105}\, \ln[16x]
 -\frac{188}{3}\, \pi\, s_{k}   \nonumber \\
 &\quad-\frac{151}{6}\, \pi\, (X_1-X_2)\, \sigma_{k}+\eta\, \Bigl( -\frac{56198689}{217728}+\frac{451}{48}\, \pi^2\Bigr)
 +\frac{541}{896}\, \eta^2-\frac{5605}{2592}\, \eta^3 \biggr]  \nonumber \\
 &\quad+x^{7/2}\, \biggl[ \Bigl(-\frac{4323559}{18144}+\frac{436705}{672}\, \eta - \frac{5575}{27}\, \eta^2\Bigr)\, s_{k} \nonumber \\
  &\quad+(X_1-X_2)\, \Bigl( -\frac{1195759}{18144}+\frac{257023}{1008}\, \eta-\frac{2903}{32}\, \eta^2 \Bigr)\, \sigma_{k} \nonumber \\
 &\quad+\pi\, \Bigl( -\frac{4415}{4032}+\frac{358675}{6048}\, \eta + \frac{91495}{1512}\, \eta^2\Bigr)\biggr]
\biggr \} \,,
\end{align}
}
where $s_k$ and $\sigma_k$ are given by
\begin{subequations}
\label{eq_sk}
\begin{eqnarray}
 s_{k}&= X_1^2\, \chi_1 \, (\vek s_1 \cdot \vek k) +X_2^2\, \chi_2 \, (\vek s_2 \cdot \vek k)\,, \\
 \sigma_{k}&= X_2\, \chi_2 \, (\vek s_2 \cdot \vek k) -X_1\, \chi_1 \, (\vek s_1 \cdot \vek k)\,.
\end{eqnarray}
\end{subequations}
We would like to stress that in the above equations we have merely replaced $\vek l$ appearing in equation~(3.16) of \cite{Bohe2013}
by  $\vek k$.
This differential equation for $x$ is also 
employed in the usual implementation of the SpinTaylorT4 approximant, as provided 
by {\scshape lalsuite} SpinTaylorT4 code, while using $\vek l$ to describe the binary orbits.

\end{document}